# Highly-efficient second and third harmonic generation in a monocrystalline lithium niobate microresonator

**Jintian Lin,[1,6] Yao Ni,[2,6] Zhenzhong Hao,[3] Jianhao Zhang,[1,5] Wenbo Mao,[3] Min Wang,[4] Wei Chu,[1] Rongbo Wu,[1,5] Zhiwei Fang,[4] Lingling Qiao,[1] Wei Fang,[2,*] Fang Bo,[3,†] and Ya Cheng[1,4,5,‡]**

[1]State Key Laboratory of High Field Laser Physics, Shanghai Institute of Optics and Fine Mechanics, Chinese Academy of Sciences, Shanghai 201800, China. [2]State Key Laboratory of Modern Optical Instrumentation, College of Optical Science and Engineering, Zhejiang University, Hangzhou 310027, China. [3]The MOE Key Laboratory of Weak Light Nonlinear Photonics, TEDA Applied Physics Institute and School of Physics, Nankai University, Tianjin 300457, China. [4]State Key Laboratory of Precision Spectroscopy, East China Normal University, Shanghai 200062, China. [5]University of Chinese Academy of Sciences, Beijing 100049, China. [6]J. L. and N. Y. contributed equally to this work.

Correspondence and requests for materials should be addressed to W. F. (wfang08@zju.edu.cn); F. B. (bofang@nankai.edu.cn); or Y. C. (ya.cheng@siom.ac.cn).

**Nonlinear optics in whispering-gallery-mode (WGM) microresonators have attracted much attention. Owing to strong confinement of the light in a small volume, a WGM microresonator can dramatically boost the strength of light field, giving rise to enhancement of the nonlinear interactions of light with the resonator material. Here, we demonstrate highly efficient second harmonic generation (SHG) and third harmonic generation (THG) in an on-chip monocrystalline lithium niobate (LN) microresonator. Benefitting from a cyclic phase matching scheme for the transverse-electric WGMs, nonlinear wavelength conversion utilizing the largest second-order nonlinear coefficient $d_{33}$, which has not been demonstrated until now despite of its obvious advantage, is successfully achieved in an X-cut LN microresonator. We obtain high conversion efficiencies in not only the SHG (3.8% mW$^{-1}$) but also the THG (0.3% mW$^{-2}$) realized through a cascaded sum-frequency process. Our results represent a major step toward the classical and quantum photonic integrated circuits.**

## Introduction

Lithium niobate (LN) is an ideal nonlinear crystalline material for photonic applications because of its broad transmission window as well as large eletro-optic and nonlinear optical coefficients[1]. In many applications, the nonlinear optical conversion efficiency plays a key role, which can reach nearly 100% in some conventional nonlinear wavelength conversion experiments given that a phase matching condition has been fulfilled[2]. The traditional approaches to achieving the phase matched second order nonlinear wavelength conversion in a bulk crystal are based on either a birefringence effect (i.e., the natural phase matching scheme)[3] or periodical poling (i.e., quasi-phase matching scheme)[4-8]. In particular, the quasi phase matching scheme provides more flexibility in terms of the wavelengths and polarization states for the involved light waves, which can give rise to more efficient nonlinear optical processes[2,4-8]. For instance, the largest nonlinear optical coefficient (i.e., $d_{33}$ = -41.7 pm/V, which is almost one order of magnitude higher than $d_{31}$ = -

4.64 pm/V and $d_{22}$ = 2.46 pm/V)[9] of LN can only be utilized in quasi-phase matched second harmonic generation (SHG).

The advent of lithium niobate on insulator (LNOI) further enables fabrication of a variety of micro- and nano-scale photonic structures with precisely controlled geometrical shapes and low surface roughness[10,11]. In particular, LN based microdisk or microring opens the possibility to achieve efficient nonlinear wavelength conversion in a compact space, in which the confinement of light by the total internal reflection at the circular boundary of the microresonator leads to a substantial enhancement of light field[12-16]. Recently, with the development of micro/nano-fabrication techniques, strong nonlinear optical effects including SHG[12,14,17-24], third harmonic generation (THG)[19,21], sum frequency generation (SFG)[18], spontaneous and stimulated Raman scattering[21,22], four-wave mixing (FWM)[24], spontaneous parametric down conversion (SPDC)[20], and optical comb generation[25] have all been observed from high-Q LN microresonators on LNOI substrates. Nonlinear parametric processes in the LN microresonators can be realized by utilizing different phase matching conditions[12,14,17-23], yet a conversion efficiency approaching 100% has not been achieved owing mainly to three difficulties, which are a large spatial overlap between the mode fields, a nearly perfect phase matching condition, and a multiple resonance among the modes participating in the nonlinear processes.

Here, we show that by use of a high quality (Q) factor X-cut LN microresonator, we are able to achieve a normalized conversion efficiency of 3.8% $mW^{-1}$ for SHG and that of 0.3% $mW^{-2}$ for

THG. The ultimate conversion efficiencies we obtain for these two cases are 3.1% and 0.2% at the pump power of 0.825 mW. This is achieved by taking the advantage of the periodical inversion of the orientation of the light polarization with respect to the direction of the optical axis of LN in an X-cut WGM microresonator. This scheme, which is termed as cyclic-quasi-phase matching (CQPM), is conceptually similar to the $\bar{4}$-quasi-phase-matched SHG in a gallium arsenide WGM microresonator[26,27]. The CQPM leads to a few unique and significant advantages. First, the CQPM remains effective in a broad spectral bandwidth, thereby allowing for various nonlinear processes to be efficiently excited at once. Second, all of the involved light waves can be transverse-electric (TE) modes with the electric field vectors lying in the plane of the microresonator, enabling to use the largest nonlinear coefficient $d_{33}$ for the nonlinear wavelength conversions. The advantages are validated by experimentally demonstrating highly efficient SHG as well as THG which results from a SFG process between the pump and its second harmonic waves. The successfully employing $d_{33}$ in monocrystalline LN WGM microresonator renders the fabrication of periodically polarized lithium niobate (PPLN) microresonators with elaborately designed domain patterns unnecessary, although such demonstration has been reported recently[28,29]. In addition, both the pump light and the second harmonic wave are resonant with the high Q modes of Q-factors of $9.61\times10^6$ at 1547.8 nm and $3.68\times10^6$ at 773.9 nm, which are among the records of the measured Q-factors of on-chip LN microresonators reported so far. The realization of high optical conversion efficiencies obtained in an on-chip device at a low pump laser power is of vital importance for photonic integrated circuit applications.

## Results

**CQPM SHG in LN microresonator**

CQPM SHG has been realized in an X-cut LN microresonator when the fundamental wave and the second harmonic wave are both TE-polarized (i.e., the electric field in the plane of the microdisk resonator). Polarization induced in *3m* point-group materials, such as LN, depends on the non-zero susceptibility $d_{ij}$ and the polarization of the associated light waves. Typical values[9] of $d_{ij}$ are $d_{22}$= 2.46 pm/V, $d_{31}$= -4.64 pm/V, and $d_{33}$= -41.7 pm/V. Note that $d_{33}$ is almost one order of magnitude higher than the other two susceptibilities. The second-order nonlinear coefficient along the periphery of the LN microdisk can be calculated as below

$$d_{\text{eff}} = -d_{22}\cos^3\theta + 3d_{31}\cos^2\theta \sin\theta + d_{33}\sin^3\theta, \qquad (1)$$

where *θ* is the angle between the wave vector **k** of the TE-polarized wave and the optic axis. The angle *θ* is in accordance with the azimuth angle in the microdisk, as indicated in Fig. 1(a). When the light waves propagate along the periphery of the microresonator, they undergo oscillating nonlinear susceptibility $d_{\text{eff}}$. Specifically, at the azimuth angle *θ* = π/2, one can have $d_{\text{eff}}$ = $d_{33}$ neglecting the reduction due to the imperfect spatial overlap between the WGMs associated with the nonlinear optical process; whereas at *θ* = 3π/2, the nonlinear susceptibility changes to $d_{\text{eff}}$ = -$d_{33}$. The sign of $d_{\text{eff}}$ inverts every half cycle in the LN microresonator as illustrated in Fig. 1(a), which is similar to the traditional quasi-phase matching realized in PPLN crystals. The magnitude of $d_{\text{eff}}$ changes periodically with respect to *θ* in X-cut LN microresonators in contrast to PPLN microresonators.

The refractive index of the fundamental wave of a TE mode also oscillates between the maximum refractive index $n_o$ (i.e., for the ordinary beam linearly polarized along $\theta = 0$) and the minimum refractive index $n_e$ (i.e., for the extraordinary beam linearly polarized along $\theta = \pi/2$) in every half cycle when propagating around the circular microresonator. Likewise, the refractive index of the second harmonic wave experiences a synchronized periodic oscillation in the X-cut LN microresonator every half cycle whereas both the $n_o$ and $n_e$ at the second harmonic wavelength are different from that of the fundamental wave owing to the dispersion. The dispersion will cause a phase mismatch in the SHG process, which, however, can now be readily compensated thanks to the inversion of the nonlinear susceptibility every half cycle, which is similar to the quasi-phase matching in a PPLN microdisk resonator with the positive direction of the optical axis being inverted every half cycle. Such a PPLN microdisk has a poling period of $2\pi R$, where $R$ is the radius of the PPLN disk, along the circumference.

To verify the effectiveness of the CQPM scheme in nonlinear optics processes, we fabricated a LN microresonator on an X-cut LNOI substrate with a thickness of 600 nm and a diameter ($2R$) of 29.92 μm as shown by its scanning electron microscope (SEM) image in Fig. 1(b). The details on the fabrication of LN microresonator can be found in **Methods**. The side view of the microdisk is presented in Fig. 1(c), showing a smooth sidewall tilted with an angle of approximately 10° with respect to the vertical direction.

The experimental setup for generating the nonlinear optical processes can be found in **Supplementary Information**. When the pump laser wavelength was set at 1547.8 nm, efficient

SHG was observed at a signal wavelength of 773.9 nm. Figure 2(a) shows the spectra of both the pump laser and the second harmonic wave. The second harmonic wave was confirmed to have a horizontal polarization (i.e, in TE mode) as well. A typical optical micrograph indicating the SHG process captured by the CCD camera is shown in the inset of Fig. 2(b), in which the second harmonic beam is clearly visible around the circumference of the microdisk. The conversion efficiency is estimated by the ratio between the power of the second harmonic and that of the pump laser. In our experiment, the power of the pump laser near the coupling region was calculated by subtracting the propagation loss of the fiber taper. The conversion efficiency grows linearly with the increasing pump power, as shown in Fig. 2(b). This is a typical feature of SHG, indicating a normalized conversion efficiency of 3.8% mW$^{-1}$ in the LN microdisk.

When the pump laser power was further increased, THG was observed. Figure 2(a) shows the spectrum of the third harmonic signal at the wavelength of 515.9 nm due to the sum-frequency process between the pump and its corresponding harmonic signal. The third harmonic wave was also detected to be of a TE mode with the electric field placed in the equator plane. The THG process was visualized using the CCD camera, as shown in the inset of Fig. 2(c). In this case, both the second and third harmonic waves were observed, but the second harmonic signal appears brighter. Figure 2(c) plots the output power of THG as a function of the pump power. The red line is a linearly fitting following the cubic pump power. The normalized conversion efficiency from the pump light to the third harmonic wave is 0.3% mW$^{-2}$. Figure 2(d) plots the output power of THG as a function of the power of SHG, which grows linearly with the increasing power of SHG following a slope of 3.79%.

## Mode Characterization and Simulation Results

To facilitate the analyses of the phase matched SHG and THG processes, the modes in the LN microresonator were characterized by measuring the transmission spectra which were recorded by the fiber taper coupling method[30]. The transmission spectra around the wavelengths of fundamental and second-harmonic waves are shown in Figs. 3(a) and 3(c), respectively. Two resonant modes denoted as $TE_{1,111}$ and $TE_{5,221}$ are identified to belong to the fundamental and second harmonic waves, respectively, which satisfy the double resonance condition required for the SHG process. The subscripts (1,111) denote the radial-mode and azimuthal-mode numbers, respectively. 111(221) is the azimuthal number $m_F$ ($m_{SHG}$) of the fundamental (second harmonic) wave. The free spectral ranges (FSRs) around the fundamental and second harmonic wavelengths are 11.63 nm and 2.85 nm, respectively. The Q factors of $TE_{1,111}$ and $TE_{5,221}$ shown respectively in Figs. 3(b) and 3(d) are determined to be $9.61 \times 10^6$ and $3.68 \times 10^6$ by Lorentz fitting. The insets in Figs. 3(b) and 3(d) are the electric field profiles of $TE_{1,111}$ and $TE_{5,221}$ modes, respectively. The simulations of mode structure and electric field profile are carried out using the finite-element method. The details of simulation can be found in **Methods**. A dense mode structure appears near the second harmonics wavelength. Since a narrowband tunable laser operated around 520 nm wavelength is not available in our lab, the mode of the third harmonic wave cannot be identified experimentally at this moment.

We have calculated the second-order nonlinear susceptibility $d_{eff}(\theta)$ along the periphery in the SHG process as a function of $\theta$, as shown in Fig. 4(a), which oscillates between -$d_{33}$ and $d_{33}$ when the angle $\theta$ varies between 0 and $\pi$. The effective refractive indices of fundamental (F) mode, second harmonic mode, and third harmonic mode as a function of the azimuth angle, are shown in Fig. 4(b). These curves oscillate as well. It is worthy of noting that the relationship between $d_{eff}$ and the propagation distance $r\theta$ of light along the circumference in an X-cut LN WGM microresonator is significantly different from that in a PPLN waveguide[31], as the effective refractive index in a PPLN can be described with a step function. Here, $r = 14.66$ μm is the horizontal position of the maximum of fundamental-wave electric field. The Fourier transform spectrum of the $d_{eff}$ with respect to $r\theta$ has two positive components in reciprocal vector $K$ space as shown by the red lines in Fig. 4(c). In a quasi-phase matched SHG, the wave vector mismatching $\Delta k$ should fulfill the following condition:

$$\Delta k = 2k_F - k_{SHG} = mK_i, \tag{2}$$

where $k_F$ and $k_{SHG}$ are the wave vectors of the pump wave and the second harmonic waves, respectively. $K_i$ is the $i$th Fourier component of $d_{eff}(r\theta)$ in the wave vector space, and $m$ is an integer. $k_F$ and $k_{SHG}$ can be derived from Fig. 4(b) showing the effective refractive indices of TE$_{1,111}$ and TE$_{5,221}$ modes with the varying $\theta$ according to $k = 2\pi n_{eff}/\lambda = m/R$, where $\lambda$ is the wavelength of wave. From Fig. 4(c) we can see that $\Delta k$ ranges from $0.406 \times 10^5$ m$^{-1}$ to $1.055 \times 10^5$ m$^{-1}$, which can be compensated in several points by $K_1 = 0.682 \times 10^5$ m$^{-1}$, one of the non-zero Fourier components of $d_{eff}(r\theta)$.

Owing to the cyclic behavior of the SHG process under the double resonance condition in the microresonator, the growth of the second harmonic signal can now be investigated in the range of 0 ~ 2π for $\theta$. Under the slowly varying amplitude approximation, the growth rate of the field amplitude $E_{SHG}$ with the varying $\theta$ can be expressed as[17,32]

$$\frac{dE_{SHG}}{ds} = \frac{id_{eff}\omega_{SHG}^2}{k_{SHG}c^2}E_F^2 e^{i\int \Delta k ds} - \frac{E_{SHG}}{2k_{SHG}} \times \frac{dk_{SHG}}{ds}, \tag{4}$$

where $ds = rd\theta$. In Eq. (4), $\omega_{SHG}$, $k_{SHG}$, $E_F$, and $\Delta k$ (= $2k_F$ - $k_{SHG}$), are angular frequency of second harmonic, wave number of second harmonic, field amplitude of the fundamental wave, and the wavevector mismatch at the azimuth angle $\theta$, respectively. The growth of the second harmonic signal is investigated by integrating Eq. (4) along the periphery over a single cycle in the steady-state regime. Figure 4(d) shows how the field amplitude of the second harmonic wave evolves in each cycle when it propagates along the periphery of the microresonator. Similar to the quasi-phase matched SHG in PPLN, the filed amplitude grows monotonically in a full cycle when the azimuth angle varies from 0 to 2π. The SHG process then repeats itself until the second harmonic signal saturates due to the depletion of the pump laser.

Likewise, we can realize a similar quasi-phase matched THG in the same LN microresonator. Details of the calculation can be found in **Methods**.

**Conclusion**

Highly efficient SHG and THG processes have been demonstrated in an on-chip X-cut LN microresonator. The successful realization of the broadband phase matching condition uniquely utilizing the largest second-order nonlinear coefficient $d_{33}$ without the need of domain engineering will make strong impact on the microresonator-based lithium niobate photonic applications.

## Methods

**Fabrication of X-cut LN microresonator**

In our experiment, the microresonator was fabricated on a commercially available X-cut LN thin film with a thicknesses of 600 nm (NANOLN, Jinan Jingzheng Electronics Co., Ltd). The LN thin film is bonded by a silica layer with a thickness of ~2 μm on an LN substrate. Details in the process flow of fabrication of the LN microresonator have been described in Ref. [12,13], which can be summarized as the following 3 steps:

(1) The sample was first immersed in water and ablated by a focused femtosecond laser beam to form a cylindrical pad with a height of ~3 μm and a diameter of ~33 μm. Here, high-precision ablation was available by tightly focusing the femtosecond laser beam to a diameter of 1 μm. The average power of the laser beam was chosen as 0.35 mW when performing ablation in the LN material, and raised to 1 mW for ablation in the silica layer.

(2) Focused ion beam was used to polish the periphery of the cylindrical pad to ensure a smooth sidewall and therefore a high $Q$ factor of the fabricated microresonator. The diameter of the pad was reduced to ~29.92 μm after the process.

(3) Chemical wet etching with diluted hydrofluoride acid (5%) was performed to partially remove the silica layer under the LN thin film into a pedestal. The chemical wet etching lasted for 6 min. Finally a freestanding LN microdisk supported by a silica pedestal was produced.

**Transmission spectra measurement**

The transmission spectrum of 1550 nm range was measured by a tunable laser (TLB-6728, New Focus Inc.) with a linewidth of 200 kHz. To avoid thermal broadening or thermal compression of modes, the power of tunable laser was set at 0.2 µW，and the scanning speed was 88 nm/s. For measuring the transmission spectra around the 775 nm wavelength, a tunable laser (TLB-6712, New Focus Inc.) was used as the light source.

**Simulations**

A finite element method (COMSOL Multiphysics, version 5.0) was used to identify the modes in the microresonator. In a microresonator with the axial symmetry, the eigen-frequencies and field profiles of the modes can be calculated by solving the two-dimensional partial differential-equations[33]. For the TE polarized waves in the principal plane (y-z plane) of LN microresonator, the waves inevitably experience the varying refractive index $n(\theta)$ along the periphery of the LN microresonator, which oscillates between the effective ordinary value $n'_o(\lambda)$ and extraordinary value $n'_e(\lambda)$ as below,

$$\frac{1}{n^2(\lambda,\theta)} = \frac{\cos^2\theta}{n'^2_o(\lambda)} + \frac{\sin^2\theta}{n'^2_e(\lambda)}. \tag{5}$$

Therefore, an average refractive index used in the simulation of mode profile, can be obtained by integrating Eq. (5) with respect of the angle $\theta$ over 0 to $2\pi$, which is given below:

$$n_{\mathrm{av}} = \left(\frac{1}{2}\left(\frac{1}{n_o^2(\lambda)} + \frac{1}{n_e^2(\lambda)}\right)\right)^{-1/2}. \quad (6)$$

Here, $n_o(\lambda)$ and $n_e(\lambda)$ are the ordinary value and extraordinary value of the refractive indices of the bulk LN[34], respectively. Modes resonant at the frequencies of the waves participating in the nonlinear optical processes can be identified based on the simulation results.

For the TE modes, the effective refractive indices were calculated by Eq. (5), which experienced a $\theta$-dependent oscillation between the effective ordinary values $n'_o(\theta)$ and extraordinary values $n'_e(\theta)$ during circulating along the periphery. The effective ordinary (extraordinary) refractive index could be estimated from the TE-polarization WGMs in a microresonator which has a geometry exactly the same as the LN microresonator used in the experiment, however with a constant ordinary (extraordinary) refractive index.

For the THG process, the third harmonic was resonant with a high-order mode $TE_{10,335}$, of which the mode spatial distribution is shown in the inset of Fig. 4(f). The third harmonic wave was generated in the SFG process. The effective refractive indices and the wavevector mismatching as functions of the azimuth angle are shown in Figs. 4(b) and 4(e). The negative Fourier components in $K$ space are also shown by the red lines in Fig. 4 (e) from which it is clearly seen that either $3K_{-1}$ or $K_{-3}$ ($= -2.046\times10^5 \mathrm{m}^{-1}$) can compensate for the phase mismatch between the pump waves and the third harmonic signal. Figure 4(f) plots how the field amplitude of the third harmonic wave evolves in each cycle when it propagates along the periphery of the microresonator. Similar to the SHG, the filed amplitude grows efficiently in a full cycle when the azimuth angle varies from 0 to $2\pi$.

# References

1  Wong, H. K. *Properties of lithium niobate* (INSPEC, The Institution of Electrical Engineers, London, 2002).

2  Boyd, R. W. *Nonlinear optics* (Academic Press, Singapore, 2008).

3  Boyd, R. W. *et al*. LiNbO$_3$: An efficient phase matchable nonlinear optical material. *Appl. Phys. Lett*. **5**, 234 (1964).

4  Feng, D. *et al*. Enhancement of second-harmonic generation in LiNbO$_3$ crystals with periodic laminar ferroelectric domains. *Appl. Phys. Lett*. **37**, 607 (1980).

5  Matsumoto, S., Lim, E. J., Hertz, H. M. & Fejer, M. M. Quasi phase-matched second harmonic generation of blue light in electrically periodically-poled lithium tantalate waveguides. *Electron. Lett*. **27**, 2040–2042 (1991)

6  Jin, H. *et al*. On-chip generation and manipulation of entangled photons based on reconfigureable lithium-niobate waveguide circuits. *Phys. Rev. Lett*. **113**, 103601 (2014).

7  Zhang, J., Chen, Y., Lu, F. & Chen, X. Flexible wavelength conversion via cascaded second order nonlinearity using broadband SHG in MgO-doped PPLN. *Opt. Express* **16**, 6957–6962 (2008).

8  Ilchenko, V. S., Savchenkov, A. A., Matsko, A. B. & Maleki, L. Nonlinear optics and crystalline whispering gallery mode cavities. *Phys. Rev. Lett.* **92**, 043903 (2004).

9  Dmitriev, V. G., Gurzadyan, G. G. & Nikogosyan, D. N. *Handbook of nonlinear optical*

*crystals* (Springer-Verlag, Berlin, 1999).

10  Poberaj, G., Hu, H., Sohler, W. & Günter, P. Lithium niobate on insulator (LNOI) for micro-photonic devices. *Laser Photon. Rev.* **6**, 488–503 (2012).

11  Boes, A. *et al*. Status and potential of lithium niobate on insulator (LNOI) for photonic integrated circuits. *Laser Photon. Rev.* **12**, 1700256 (2018).

12  Lin, J. *et al*. Second harmonic generation in a high-Q lithium niobate microresonator fabricated by femtosecond laser micromachining. Preprint at http://arxiv.org/abs/1405.6473 (2014).

13  Lin, J. *et al*. Fabrication of high-Q lithium niobate microresonators using femtosecond laser micromachining. *Sci. Rep.* **5**, 8072 (2015).

14  Wang, C. *et al*. Integrated high quality factor lithium niobate microdisk resonators. *Opt. Express* **22**, 30924–30933 (2014).

15  Wang, J. *et al*. High-Q lithium niobate microdisk resonators on a chip for efficient electro-optic modulation. *Opt. Express* **23**, 23072–23078 (2015).

16  Wang, L. *et al*. High-Q chaotic lithium niobate microdisk cavity. *Opt. Lett*. **43**, 2917–2920 (2018).

17  Lin, J. *et al*. Phase-matched second-harmonic generation in an on-chip microresonator. *Phys. Rev. Appl.* **6**, 014002 (2016).

18  Hao, Z. *et al*. Sum-frequency generation in on-chip lithium niobate microdisk resonators. *Photon. Res*. **6**, 623–628 (2017).

19  Liu, S., Zheng, Y. & Chen, X. Cascading second-order nonlinear processes in a lithium niobate-on-insulator microdisk. *Opt. Lett*. **42**, 3626–3628 (2017).

20  Luo, R. *et al*. On-chip second-harmonic generation and broadband parametric down-conversion in a lithium niobate microresonator. *Opt. Express* **25**, 24531–24539 (2017).

21  Wolf, R., Breunig, I., Zappe, H. & Buse, K. Cascaded second-order optical nonlinearities in on-chip micro rings. *Opt. Express* **25**, 29927–29933 (2017).

22  Wu, R. *et al*. Lithium niobate micro-disk resonators of quality factors above 10^7. Preprint at https://arxiv.org/abs/1806.00099 (2018).

23  Moore, J. *et al*. Efficient second harmonic generation in lithium niobate on insulator. *Proc. Conf. Lasers and Electro-Optics (CLEO)*, pp. STh3P.1 (2016).

24  Wang, M. *et al*. Simultaneous generation of four-wave mixing and stimulated Raman scattering in coupled lithium niobate microdisks mediated by second harmonic generation. Preprint at https://arxiv.org/abs/1803.11341 (2018).

25  Wang, C. *et al*. On-chip Kerr frequency comb generation in lithium niobate microresonators. *Proc. Conf. Lasers and Electro-Optics (CLEO)*, pp. SW4M.3 (2018).

26  Kuo, P. S., Bravo-Abad, J. & Solomon, G. S. Second-harmonic generation using $\bar{4}$-quasi-phasematching in a GaAs whispering-gallery-mode microcavity. *Nat. Commun.* **5**, 3109 (2014).

27  Kuo, P. S., Fang, W. & Solomon, G. S. $\bar{4}$-quasi-phase-matched interactions in GaAs microdisk cavities. *Opt. Lett*. **34**, 3580–3582 (2009).

28  Hao, Z. *et al*. Periodically poled lithium niobate whispering gallery mode microcavities on a chip. *Sci. China-Phys. Mech. & Astron*. **61**, 114211 (2018).

29  Wolf, R. *et al*. Quasi-phase-matched nonlinear optical frequency conversion in on-chip whispering galleries. *Optica* **5**, 872–875 (2018).

30  Knight, J. C., Cheung, G., Jacques, F. & Birks, T. A. Phase-matched excitation of whispering-gallery-mode resonances by a fiber taper. *Opt. Lett.* **22**, 1129–1131 (1997).

31  Haertle, D. Domain patterns for quasi-phase matching in whispering-gallery modes. *J. Opt*. **12**, 035202 (2010).

32  Lin, G., Fürst, J. U., Strekalov, D. V. & Yu, N. Wide-range cyslic phase matching and second harmonic generation in whispering gallery resonayors. *Appl. Phys. Lett.* **103**, 181107 (2013).

33  Oxborrow, M. Traceable 2-D finite-element simulation of the whispering-gallery modes of axisymmetric electromagnetic resonators. *IEEE Trans. Microw. Theory Tech.* **55**, 1209–1218 (2007).

34  Hobden, M. V. & Warner, J. The temperature dependence of the refractive indices of pure lithium niobate. *Phys. Lett*. **22**, 243–244 (1966).

## Acknowledgements

The work is supported by National Basic Research Program of China (Grant No. 2014CB921303), National Natural Science Foundation of China (Grant Nos. 11734009, 61590934, 61635009, 61327902, 61505231, 11604351, 11674340, 61575211, 61675220, 11674181), the Strategic Priority Research Program of Chinese Academy of Sciences (Grant No. XDB16000000), Key Research Program of Frontier Sciences, Chinese Academy of Sciences (Grant No. QYZDJ-SSW-SLH010), the Project of Shanghai Committee of Science and Technology (Grant 17JC1400400), Shanghai Rising-Star Program (Grant No. 17QA1404600), and the Fundamental Research Funds for the Central Universities.

**Figure Captions:**

Fig. 1 (a) The variations of effective second-order nonlinear optical coefficient and polarization state of TE-polarized wave are schematically illustrated in an X-cut LN microresonator. (b) SEM image of the microresonator, Inset: the side-view SEM image of the microresonator.

Fig. 2 (a) Spectra of the pump light, the second harmonic wave, and the third harmonic wave. (b) The conversion efficiency of SHG as a function of the power of the pump light, (c) The power of THG as a function of the cubic power of the pump light. Inset: top-view optical micrograph of the THG from the microresonator. (d) The power of the THG as a function of the power of second harmonic.

Fig. 3 (a) Transmission spectrum around pump wavelength recorded by coupling the fiber taper with the microresonator. (b) Lorentzian fit (red line) of measured spectrum (black line) around $TE_{1,111}$ mode, showing a Q factor of $9.61 \times 10^6$. (c) Transmission spectrum around second-hamonic wavelength. (d) Lorentzian fit (red solid line) of the measured spectrum (black line) around $TE_{5,221}$ mode, showing a Q factor of $3.68 \times 10^6$.

Fig. 4 (a) Effective second-order nonlinear coefficient at every azimuth angle. Numerically calculated (b) Effective refractive indices versus the azimuth angle, (c) Left: phase mismatching $\Delta k$ versus the azimuth angle, right: positive Fourier components in the wavevector space for the Fourier transform of (a), (d) The electric field amplitude of second harmonic signal versus the azimuth angle along the microresonator periphery, (e) Left: mismatch between wave vectors $\Delta k'\ (= k_\text{F} + k_\text{SHG} - k_\text{THG})$, right: negative Fourier components in the wavevector space for the Fourier transform of (a), (f) The electric field amplitude of third harmonic signal versus the azimuth angle along the microresonator periphery.

Fig. 1

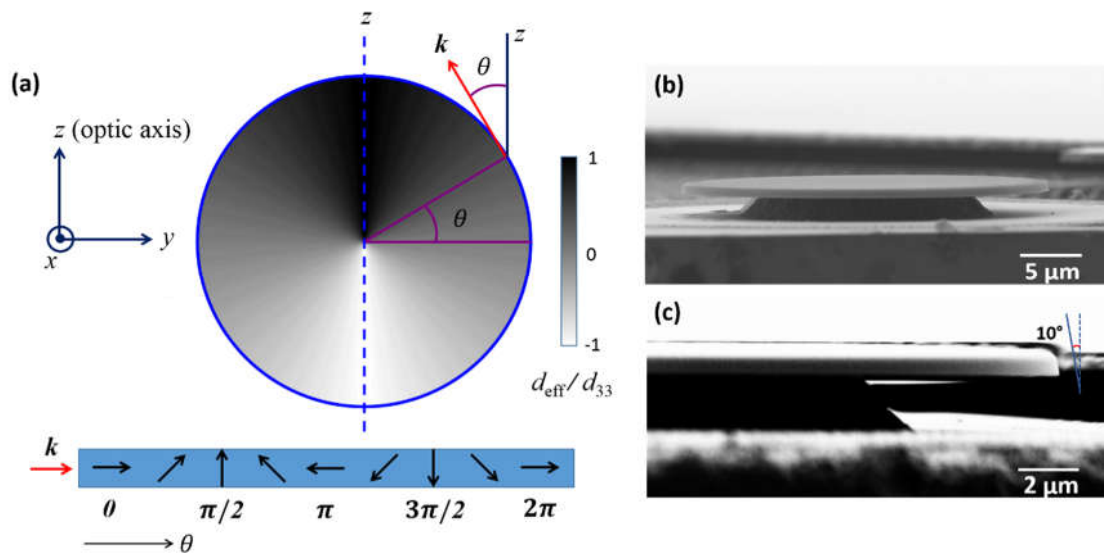

Fig. 2

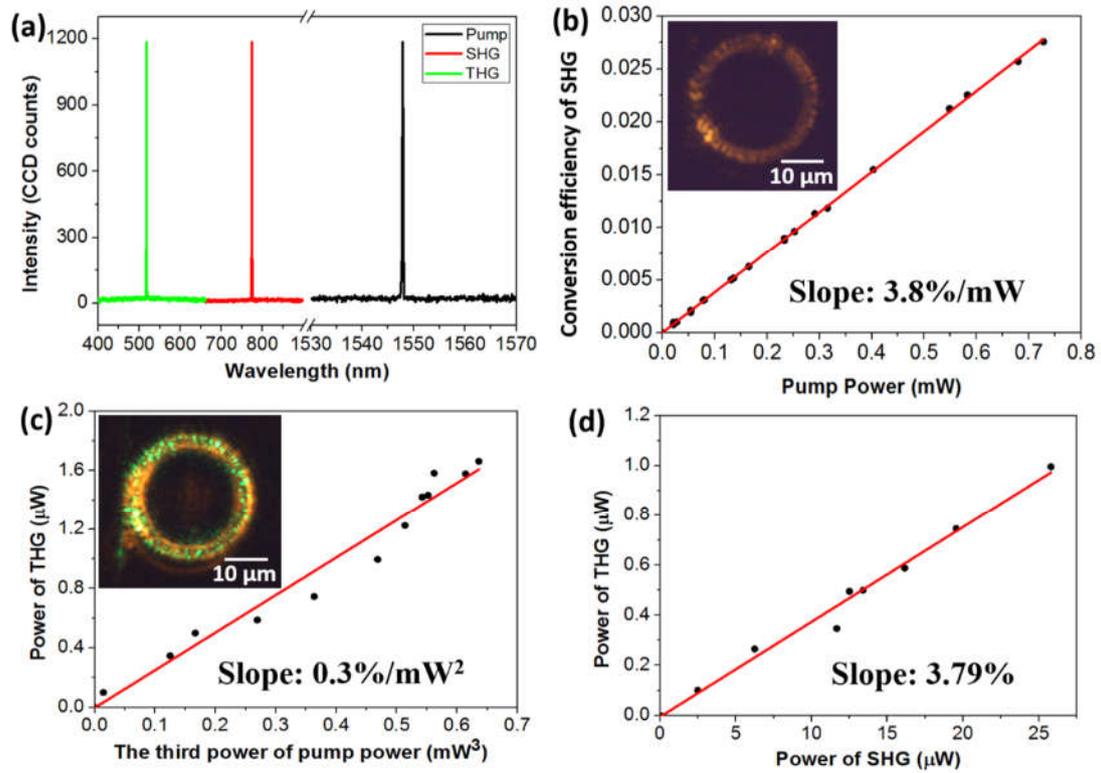

Fig. 3

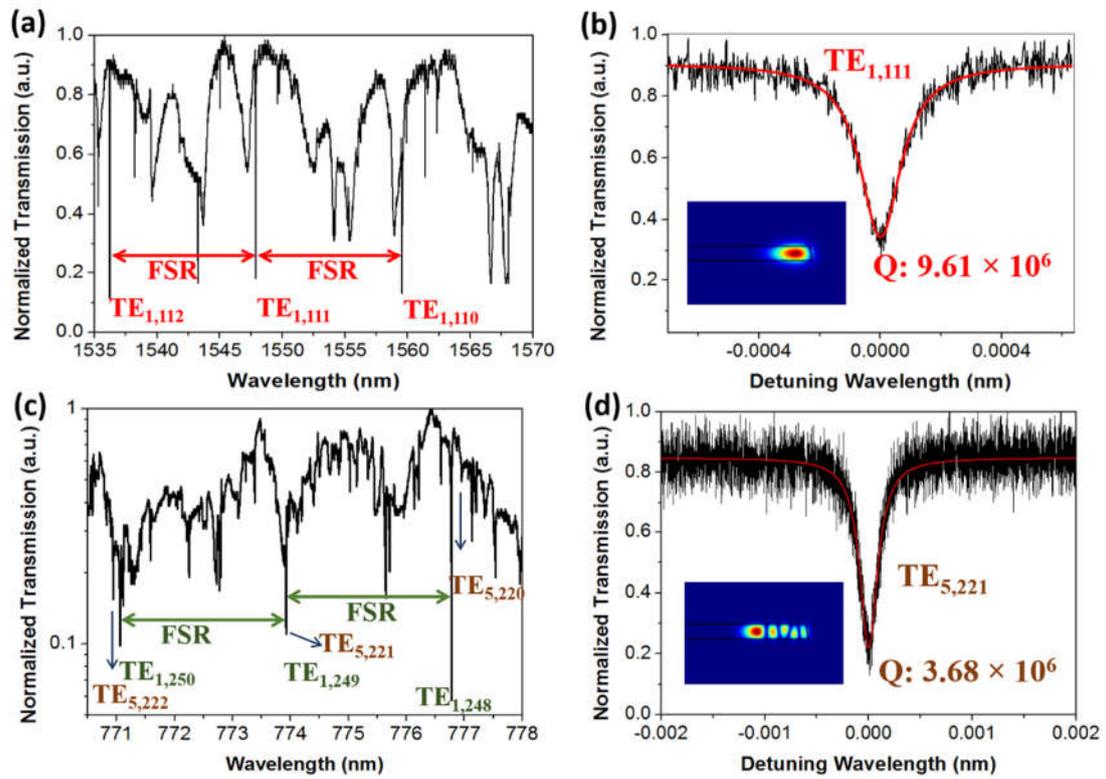

Fig. 4

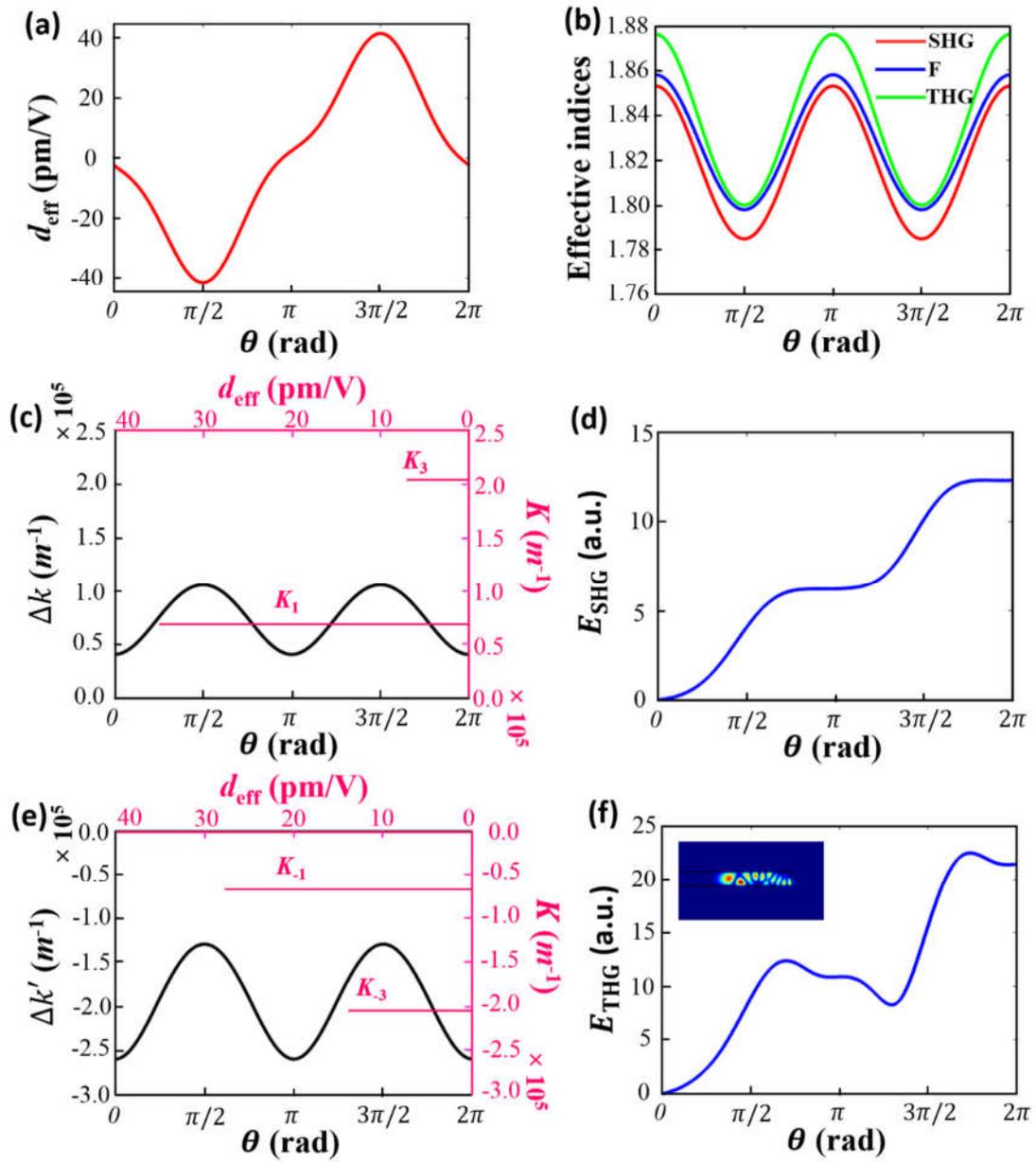